%% file: bias.tex
\documentclass[useAMS,usenatbib]{mn2e}
\usepackage{graphicx}
\usepackage{epsfig} 
\usepackage{multirow} 
\usepackage{subfigure}

\def\gsim{\;\rlap{\lower 2.5pt \hbox{$\sim$}}\raise 1.5pt\hbox{$>$}\;} 
\def\lsim{\;\rlap{\lower 2.5pt \hbox{$\sim$}}\raise 1.5pt\hbox{$<$}\;}

\include{mydefs}

\title[Halo Bias and Mass Function in Non-Markovian Excursion Set Theory]
{The Bias and Mass Function of Dark Matter Haloes in Non-Markovian
  Extension of the Excursion Set Theory}

\author[Chung-Pei Ma, Michele Maggiore, Antonio Riotto, Jun Zhang]
{Chung-Pei Ma$^{1}$, Michele Maggiore$^{2}$, Antonio Riotto$^{3,4}$, Jun Zhang$^{5}$ \\
\\
$^{1}$Department of Astronomy, University of California, Berkeley, CA 94720, USA \\
$^{2}$D\'epartement de Physique Th\'eorique, Universit\'e de 
	     Gen\`eve, CH-1211 Geneva, Switzerland\\
$^{3}$CERN, PH-TH Division, CH-1211, Gen\`eve 23,  Switzerland\\
$^{4}$INFN, Sezione di Padova, Via Marzolo 8,
I-35131 Padua, Italy\\
$^{5}$Texas Cosmology Center, University of Texas, Austin, TX 78712 USA \\ 
}

\begin{document}

\pagerange{ 
\pageref{firstpage}-- 
\pageref{lastpage}} 

\maketitle

\label{firstpage} 
\begin{abstract}
  The excursion set theory based on spherical or ellipsoidal gravitational
  collapse provides an elegant analytic framework for calculating the mass
  function and the large-scale bias of dark matter haloes.  This theory
  assumes that the perturbed density field evolves stochastically with the
  smoothing scale and exhibits Markovian random walks in the presence of a
  density barrier.  Here we derive an analytic expression for the halo bias
  in a new theoretical model that incorporates non-Markovian extension of
  the excursion set theory with a stochastic barrier.  This model allows us
  to handle non-Markovian random walks and to calculate perturbativly these
  corrections to the standard Markovian predictions for the halo mass
  function and halo bias.  Our model contains only two parameters:
  $\kappa$, which parameterizes the degree of non-Markovianity and whose
  exact value depends on the shape of the filter function used to smooth
  the density field, and $a$, which parameterizes the degree of
  stochasticity of the barrier.  Appropriate choices of
  $\kappa$ and $a$ in our new model can lead to a closer match to both the
  halo mass function and halo bias in the latest $N$-body simulations than
  the standard excursion set theory.
 \end{abstract}

\section{Introduction}\label{intro}

Dark matter haloes typically form at sites of high density peaks.  The
spatial distribution of dark matter haloes is therefore a biased tracer of
the underlying mass distribution.  A standard way to quantify this
difference between haloes and mass is to use a halo bias parameter $b_h$, which
can be defined as the ratio of the overdensity of haloes to mass, or as the
square root of the ratio of the two-point correlation function (or power
spectrum) of haloes to mass.  

Like the halo mass function, analytic expressions for the halo bias can be
obtained from the excursion set theory \citep{Bond} based on the spherical
gravitational collapse model \citep{CK89,MW96}.  In the excursion set
theory, the density perturbation evolves stochastically with the smoothing
scale, and the problem of computing the probability of halo formation is
mapped into the first-passage time problem in the presence of a (constant)
barrier.  The approach to the clustering evolution is based on a
generalization of the peak-background split scheme \citep{Bardeen}, which
basically consists in splitting the mass perturbations into a fine-grained
(peak) component filtered on a scale $R$ and a coarse-grained (background)
component filtered on a scale $R_0\gg R$.  The underlying idea is to
ascribe the collapse of objects on small scales to the high frequency modes
of the density fields, while the action of large-scale structures of these
non-linear condensations is due to a shift of the local background density.

Comparison with $N$-body simulations finds that the spherical collapse
model underpredicts the halo bias for low mass halos \citep{jing98, ST99}.
The discrepancy reaches a factor of $\sim 2$ at $M\sim 0.01 M_*$, where
$M_*$ is the characteristic nonlinear mass scale (defined by
$\sigma(M_*)=1$ where $\sigma^2(M)$ is the variance of the density field in
a volume of radius $R$ containing the mass $M$).  \cite{SMT01} obtained an
improved formula for the halo bias by using a moving barrier whose
scale-dependent shape is motivated by the ellipsoidal gravitational
collapse model.  Compared to the spherical collapse model, this formula
predicts a lower bias at the high mass end and a higher bias at the low
mass end (see Fig.~1 below).  The resulting bias is shown to be too high
at the low mass end by $\sim 20$\% compared with simulation results.
Further modifications have been introduced that either used the functional
form of \cite{ST99} or \cite{SMT01} with new fitting parameters (e.g.,
\citealt{tinker05}), or proposed new fitting forms altogether (e.g.,
\citealt{SW04, pillepich10,tinker10}).

Our goal in this paper is not to improve on the accuracy of the fits to the
halo bias, but rather to gain deeper theoretical insight 
by deriving an analytical expression for the halo bias using a new model.
This model modifies the excursion set theory by incorporating non-Markovian
random walks in the presence of a stochastic barrier.  It is based
on a path integral formulation introduced in \cite{MR1, MR2}, which
provides an analytic framework for calculating perturbatively the
non-Markovian corrections to the standard version of the excursion set
theory.  Mathematically, the non-Markovianity in the theory is related to
the choice of the filter function necessary to smooth out the density
contrast. As soon as the filter function is different from a step (tophat)
function in momentum space, the excursion of the smoothed density contrast
is non-Markovian, namely, every step depends on the previous ones and the
random walk acquires memory. As the computation of the bias parameter $b_h$
amounts to computing the first crossing rate with a non-trivial initial
condition at a large, but not infinite, radius, the non-Markovianity makes
the calculation much harder than the Markovian case.

Furthermore, the critical value for collapse in our model is itself assumed
to be a stochastic variable, whose scatter reflects a number of complicated
aspects of the underlying dynamics.   The gravitational collapse of haloes is a 
complex dynamical phenomenon, and modeling it as spherical, or even as
ellipsoidal, is a significant oversimplification.  In addition, the very
definition of what is a dark matter halo, both in simulations and
observationally, is a non-trivial problem.  \cite{MR2} proposed
that some of the physical complications inherent to a realistic description
of halo formation can be included in the excursion set theory framework, at
least at an effective level, by taking into account that the critical value
for collapse is itself a stochastic variable.

In Section~2 we review briefly the derivation for the halo mass function in
the Markovian excursion set theory (Sec 2.1) and the path integral approach
used to introduce non-Markovian terms (Sec 2.2) and stochastic barriers
(Sec 2.3) into the theory.  Section~3 is devoted to the discussion of halo
bias, including a review of the standard derivation in the Markovian case
(Sec 3.1), and a summary of our new derivation in the non-Markovian model
(Sec 3.2 and 3.3).  The details of how the halo bias is calculated from the
conditional probability for two barrier crossings in the new model is
provided in the Appendix.  In Section~4 we compare the predictions for the
halo bias and mass function in our new model with those from the Markovian
model (both spherical and ellipsoidal collapse) and $N$-body simulations.

\section{Non-Markovian Extension and Stochastic Barrier}

In this section we review the main points of the excursion set theory
\citep{Bond} and then summarize how to introduce non-Markovian terms
in the presence of a stochastic barrier \citep{MR1, MR2}.

\subsection{Brief review of the excursion set theory}

The basic variable is the smoothed density contrast,
\be\label{dfilter}
\d({\bf x},R) =\int d^3x'\,  W(|{\bf x}-{\bf x}'|,R)\, \d({\bf x}')\, ,
\ee
where $\delta({\bf x}) = \rho({\bf x})/\bar\rho - 1$ is the density
contrast about the mean mass density $\bar\rho$ of the universe, $W(|{\bf
  x}-{\bf x}'|,R)$ is the filter function, and $R$ is the smoothing scale.
We are interested in the evolution of $\d({\bf x},R)$ with smoothing scale
$R$ at a fixed point ${\bf x}$ in space, so we suppress the argument ${\bf
  x}$ from this point on. It is also convenient to use, instead of $R$, the
variance $S$ of the smoothed density field defined by
\begin{equation}
   S(R) \equiv \sigma^2(R) = \int \frac{d^3k}{(2\pi)^3}\, P(k) \tilde{W}^2(k, R) \,,
\label{S}
\end{equation}
where $P(k)$ is the power spectrum of the matter density fluctuations in
the cosmological model under consideration, and $\tilde{W}$ is the Fourier
transform of the filter function $W$.  A smoothing radius $R=\infty$
corresponds to $S=0$ and, in hierarchical models of structure formation
such as the $\Lambda$CDM model, $S$ is a monotonically decreasing function
of $R$.  We can therefore use $S$ and $R$ interchangeably and denote our
basic variable by $\d(S)$.

If the filter function is taken to be a tophat in momentum space, $\d(S)$
then satisfies a simple Langevin equation with $S$ playing the role of a
``pseudo-time'' \citep{Bond} 
\be\label{Langevin1} 
  \frac{\pa\d(S)}{\pa S} = \eta(S)\, , 
\ee
where $\eta(S)$ represents a stochastic ``pseudo-force'' whose two-point
correlation statistic obeys a Dirac-delta function: \be\label{Langevin2}
\langle \eta(S_1)\eta(S_2)\rangle =\d_D (S_1-S_2)\, .  \ee It then follows
that the function $\Pi(\d_0;\d;S)$, which gives the probability density of
reaching a value $\d$ at ``time'' $S$ starting at a value $\d_0$ at $S=0$,
satisfies the Fokker-Planck equation
\be\label{FPdS} 
  \frac{\pa\Pi}{\pa S}=\frac{1}{2}\, \frac{\pa^2\Pi}{\pa \d^2}\, .  
\ee
In \cite{Bond} this equation was supplemented by the boundary condition
\be\label{bc} \left.\Pi (\d,S)\right|_{\d=\d_c}=0 \ee to eliminate the
trajectories that have reached the critical value $\d_c$ for collapse.

The corresponding solution of the Fokker-Planck equation is 
\be\label{PiChandra}
\Pi (\d_0;\d;S)=\frac{1}{\sqrt{2\pi S}}\,
\[  e^{-(\d-\d_0)^2/(2S)}- e^{-(2\d_c-\d_0-\d)^2/(2S)} \] \,.
\ee
The probability ${\cal F}(S) dS$ of first crossing the threshold density
$\delta_c$ between ``time'' $S$ and $S+dS$ is then given by
\begin{equation}
{\cal F}(S)  = -\int_{-\infty}^{\d_c}d\d\, \frac{\pa\Pi}{\pa S}
     = \frac{1}{\sqrt{2\pi}}\, \frac{\delta_c}{S^{3/2}} e^{-\delta_c^2/(2S)}\,,
\label{calFmarkov}
\end{equation}
where we have set $\d_0=0$.  The number density of virialized objects with
mass between $M$ and $M+dM$ is related to the first crossing probability
between $S$ and $S+dS$ by
\begin{eqnarray}
   \frac{dn(M)}{dM}  dM & = &\frac{\bar{\rho}}{M} {\cal F}(S) dS \nonumber \\
  &=& \sqrt{\frac{2}{\pi}}\,\frac{\d_c}{\s}\, e^{-\d_c^2/(2\s^2)}
\, \frac{\bar{\rho}}{M^2} \frac{d\ln\s^{-1}}{d\ln M} dM \nonumber \\
  &\equiv & f(\nu)\, \frac{\bar{\rho}}{M^2} \frac{d\ln\s^{-1}}{d\ln M} dM \, ,
\label{massfunction}
\end{eqnarray}
where $\sigma=S^{1/2}$ is defined in equation~(\ref{S}) and $\nu \equiv
\delta_c/\sigma$.  This expression reproduces the mass function of
\cite{PS}, including the correct overall normalization that had to be
adjusted by hand in \cite{PS}.  We will use the dimensionless function
$f(\nu)$ defined in equation~(\ref{massfunction}) to denote the halo mass
function below.


\subsection{Non-Markovian extension}

As already discussed in \cite{Bond}, a difficulty of the excursion set
approach in Sec~2.1 is that an unambiguous relation between the smoothing
radius $R$ and the mass $M$ of the corresponding collapsed halo only exists
when the filter function is a tophat in coordinate space: $M(R)=(4/3)\pi
R^3 \rho$.  For all other filter functions (e.g., tophat in momentum space,
Gaussian), it is impossible to associate a well-defined mass $M(R)$ (see
also the recent review \citealt{zentner07}).  More importantly, $\d(S)$
obeys a Langevin equation with a Dirac delta noise as in
\eqs{Langevin1}{Langevin2} {\it only} when the filter function is a tophat
in momentum space.  Otherwise, the evolution of $\d$ with the smoothing
scale becomes non-Markovian, and the distribution function $\Pi
(\d_0;\d;S)$ of the trajectories no longer obeys the Fokker-Planck
equation, nor any local generalization of it.  In this case, $\Pi
(\d_0;\d;S)$ obeys a complicated equation that is non-local with respect to
the variable $S$ \citep{MR1}.

To deal with this problem, \cite{MR1} proposed a ``microscopic'' approach,
in which one computes the probability associated with each trajectory
$\d(S)$, and sums over all relevant trajectories.  As with any path
integral formulation, it is convenient to discretize the time variable and
to take the continuum limit at the end. Therefore we discretize the
interval $[0,S]$ in steps $\D S=\eps$, so $S_k=k\eps$ with $k=1,\ldots n$,
and $S_n\equiv S$, and a trajectory is defined by the collection of values
$\{\delta_1,\ldots ,\delta_n\}$, such that $\delta(S_k)=\delta_k$.  All
trajectories start at a value $\d_0$ at time $S=0$.

The basic quantity in this approach is the probability density in the space
of trajectories, defined as
\be\label{defW} 
   W(\delta_0;\delta_1,\ldots ,\delta_n;S_n)\equiv
   \langle \d_D (\delta(S_1)-\delta_1)\ldots \d_D
    (\delta(S_n)-\delta_n)\rangle
\ee
where $\d_D$ denotes the Dirac delta function.  In terms of $W$ we define
\be
\label{defPi} 
    \Pi_{\eps} (\delta_0;\delta_n;S_n)
  \equiv\int_{-\infty}^{\delta_c} d\delta_1\ldots d\delta_{n-1}\,
    W(\delta_0;\delta_1,\ldots ,\delta_n;S_n)
\ee 
where $S_n=n\eps$, and
$\Pi_{\eps} (\delta_0;\d;S)$ is the probability density of arriving at the
``position" $\d$ in a ``time'' $S$, starting from $\delta_0$ at time
$S_0=0$, through trajectories that never exceeded $\delta_c$. The problem
of computing the distribution function of excursion set theory is therefore
mapped into the computation of a path integral with a boundary at
$\d=\d_c$.

The probability density $W$ can be computed in terms of the connected
correlators of the theory.  When the density field $\delta$ is a Gaussian
random variable, only the two-point connected function is non-zero, and one
finds
\bees\label{WnNG0}
&&W(\delta_0;\delta_1,\ldots ,\delta_n;S_n)=\\
&&
\inT\frac{d\lambda_1}{2\pi}\ldots\frac{d\lambda_n}{2\pi}\,
e^{ i\sum_{i=1}^n\lambda_i\delta_i
-\frac{1}{2}\sum_{i,j=1}^n\lambda_i\lambda_j
\langle\delta_i\delta_j\rangle_c}
\, ,\nn
\ees
%
and $\delta_i\equiv\delta(S_i)$.  We will restrict the discussion here to
the Gaussian case since higher-order connected correlators must be included
in the non-Gaussian case \citep{MR3,MR4}.

Consider first the case of a tophat filter in momentum space, so the
evolution of $\d(S)$ is Markovian and obeys
\eqs{Langevin1}{Langevin2}. Then one can show that the connected two-point
correlator is given by
\be
\langle\delta(S_i)\delta(S_j)\rangle_c={\rm min}(S_i,S_j)\,,
\ee 
and the integrals over $d\lambda_1,\ldots ,d\lambda_n$ in
\eq{WnNG0} can be performed explicitly to give
\be\label{W} W^{\rm
  gm}(\delta_0;\delta_1,\ldots ,\delta_n;S_n)=\frac{1}{(2\pi\eps)^{n/2}}\,
e^{-\frac{1}{2\eps}\, \sum_{i=0}^{n-1} (\delta_{i+1}-\delta_i)^2}
\ee
where the superscript ``gm'' stands for ``Gaussian and Markovian.''
Inserting this expression into \eq{defPi} it can be shown \citep{MR1} that,
in the continuum limit, the corresponding distribution function $\Pi^{\rm
  gm}_{\eps=0} (\delta_0;\delta;S)$ satisfies the Fokker-Planck equation
(\ref{FPdS}) as well as the boundary condition (\ref{bc}), and therefore we
recover the standard result (\ref{PiChandra}) of excursion set theory.

The interesting case is to generalize the computation above to filter
functions different from the conventional tophat in momentum space.  The
two-point correlator depends on the filter function.  For a Gaussian filter
and a tophat filter in coordinate space, for instance, we find
\be\label{rever} 
   \langle\delta (S_i)\delta (S_j)\rangle = {\rm
  min}(S_i,S_j) + \D(S_i,S_j)\, , 
\ee 
where $\D(S_i,S_j)=\D(S_j,S_i)$ and, for $S_i\leq S_j$, the function
$\D(S_i,S_j)$ is well approximated by
\be \label{approxDelta}
\D(S_i,S_j)\simeq \kappa\, \frac{S_i(S_j-S_i)}{S_j}\, ,
\ee
with $\kappa\approx 0.35$ for a Gaussian filter and $\kappa\approx 0.44$
for a tophat filter in coordinate space. The parameter $\kappa$ gives a
measure of the non-Markovianity of the stochastic process, and the
computation of the distribution function $\Pi (\delta_0;\delta;S)$ can be
performed order by order in $\kappa$. The technique necessary for
evaluating the path integral in \eq{defPi} to first order in $\kappa$ has
been developed in \cite{MR1}, and will be further discussed below.  To
first order in the non-Markovian corrections, the resulting first-crossing
rate becomes
\be\label{Ffinal}
{\cal F}(S)=\frac{1-\kappa}{\sqrt{2\pi}}\, \frac{\delta_c}{S^{3/2}}
e^{-\delta_c^2/(2S)}+ \frac{\kappa}{2\sqrt{2\pi}}
\frac{\delta_c}{S^{3/2}}\G\(0,\frac{\delta_c^2}{2S}\)\, , 
\ee
where $\G(0,z)$ is the incomplete Gamma function.  For $\kappa=0$ one
recovers the Markovian result in equation~(\ref{calFmarkov}).  The halo
mass function is then obtained by substituting this expression for ${\cal
  F}$ into equation~(\ref{massfunction}).

\subsection{Stochastic barrier}

The constant barrier $\d_c\simeq 1.686$ in the spherical collapse model
is a significant oversimplification of the complex dynamics leading to halo
formation and growth.  Such a model can be improved in various ways.  For
instance, the excursion set theory results for the mass function 
have been shown to match more closely those from $N$-body
simulations by considering a moving barrier whose shape is motivated by the
ellipsoidal collapse model \citep{ST99,SMT01,ST02,DSMR}.  The equations are
summarized in Table~1.
%
%
%
The parameters $a, b, c$ are fixed by fit to $N$-body simulations, while
$A$ is fixed by the normalization condition on the halo mass function.  As
already remarked in \cite{SMT01}, these expressions can be obtained from a
barrier shape that is virtually identical to the ellipsoidal collapse
barrier, except for the factor of $a$, which is not a consequence of the
ellipsoidal collapse model. In fact, the ellipsoidal collapse model reduces
to the spherical collapse model in the large mass limit. In this limit the
mass function is determined by the slope of the exponential factor, so even
in an ellipsoidal collapse model we must have $a=1$, as in the spherical
model. However, numerical simulations show that $a<1$ and its precise value
also depends on the details of the algorithm used for identifying halos in
the simulation, e.g., the link length in a friends-of-friends (FOF) halo
finder, or the critical overdensity in a spherical density (SO) finder.

A physical understanding of the parameter $a$ is given by a second
independent improvement of the spherical collapse model, the diffusing
barrier model proposed in \cite{MR2}.  These authors suggested that at
least some of the physical complications inherent to a realistic
description of halo formation, which involves a mixture of smooth
accretion, violent encounters and fragmentations, can be included in the
excursion set theory framework by assuming that the critical value for
collapse is itself a stochastic variable, whose scatter reflects a number
of complicated aspects of the underlying dynamics. In the simple example of
a barrier performing a random walk with diffusion coefficient $D_B$ around
the spherical collapse barrier, one finds indeed a mass function in which
$\d_c$ is effectively replaced by $a^{1/2}\d_c$, with $a=1/(1+D_B)$, while
at the same time $\kappa$ is replaced by $a\kappa$ (see Table~1).

\begin{table*}
\centering
\caption{Summary of Mass Function and Halo Bias Predicted by Various Analytic Models}
\begin{tabular}{lccl}
  \hline
   Model &   Mass Function $f(\nu)$  &  Halo Bias $b_h(\nu)$ &  Parameters\\
      \hline

   Spherical Collapse  & $\sqrt{\frac{2}{\pi}} \nu \exp\left(-\frac{\nu^2}{2}\right)$ &
           $1 + \frac{\nu^2-1}{\delta_c}$ &  $\delta_c=1.686$ \\
    \\
   Ellipsoidal Collapse & $A \sqrt{\frac{2}{\pi}} \sqrt{a}\nu 
         \exp\left(-\frac{a\nu^2}{2}\right) \left[ 1+(a\nu^2)^q \right]$  &
           $1 + \frac{1}{\sqrt{a}\delta_c} \left[ \sqrt{a}(a\nu^2)
            + \sqrt{a} b (a\nu^2)^{1-c} \right. $
        &  $A=0.322, q=-0.3$ \\
   &&  $\left. -\frac{(a\nu^2)^c}{(a\nu^2)^c + b(1-c)(1-c/2)}\right]$ &
       $a=0.707, b=0.5, c=0.6$ \\
     \\
  Non-Markovian & $\sqrt{\frac{2}{\pi}} \left[(1-\kappa)\nu 
          \exp\left(-\frac{\nu^2}{2}\right) + \kappa \frac{\nu}{2} 
       \Gamma\left(0,\frac{\nu^2}{2}\right)   \right]$ &
            $1+ \frac{1}{\delta_c \left[ 1-\kappa+ \frac{\kappa}{2} e^{\nu^2/2}\Gamma(0,\nu^2/2) \right] }
        \left\{ \nu^2-1 \right.$ &
         $\kappa=0$ for tophat-$k$ filter \\
   && $ \left. +\frac{\kappa}{2} \left[2- \exp\left(\frac{\nu^2}{2}\right)\Gamma\left(
        0,\frac{\nu^2}{2}\right) \right] \right\}$ & 
         $\kappa=0.35$ for Gaussian \\
    &&& $\kappa=0.44$ for tophat-$x$ \\
  \\
  Non-Markovian &
        $\kappa \rightarrow a\kappa, \quad \nu \rightarrow \sqrt{a}\nu$ &
        $\kappa \rightarrow a\kappa, \quad \nu \rightarrow \sqrt{a}\nu, \quad
        \delta_c \rightarrow \sqrt{a} \delta_c $    & 
        $a=\frac{1}{1+D_B}$ \\
  $+$ Stochastic Barrier &  && $D_B=$ diffusion coefficient\\
 \hline
\end{tabular}
\end{table*}


\section{Halo Bias}

We now apply the technique in Section~2 to the computation of the halo
bias, including the non-Markovian corrections with stochastic barriers. We
sketch here the main steps of the computations, leaving the details to the
Appendix.

\subsection{Conditional probability: the Markovian case}

To compute the bias, we need the probability of forming a halo of mass $M$,
corresponding to a smoothing radius $R$, under the condition that the
smoothed density contrast on a much larger scale $R_m$ has a specified
value $\d_m=\delta(R_m)$.  We use ${\cal F}(S_n|\d_m,S_m)$ to denote the
conditional first-crossing rate.  This is the rate at which trajectories
first cross the barrier at $\d=\d_c$ at time $S_n$, under the condition
that they passed through the point $\d=\d_m$ at an earlier time $S_m$.  We
also use the notation ${\cal F}(S_n|0)\equiv {\cal F}(S_n|\d_m=0,S_m=0)$,
so ${\cal F}(S_n|0)$ is the first-crossing rate when the
density approaches the cosmic mean value on very large scales.

The halo overdensity in Lagrangian space is given by
(\citealt{Kaiser84,EFWD, CK89,MW96}; see also \citealt{zentner07} for a
review)
\be 
   1+\d_{\rm halo}^L=\frac{{\cal F}(S_n|\d_m,S_m)}{{\cal
    F}(S_n|0)}\, .  
\ee 
In a sufficiently large region, we have $S_m\ll S_n\equiv S$ and $\d_m\ll
\d_c$.  Then, using the first crossing rate of excursion set theory and
retaining only the term linear in $\d_m$, we obtain
\be 
  \d_{\rm halo}^L=\frac{\nu^2-1}{\d_c}\,\d_m\, ,
\ee 
where $\nu =\d_c/\s$. After mapping to Eulerian space, one finds $\d_{\rm
  halo} \approx 1 + \d_{\rm halo}^L$ in the limit of small overdensity
$\d_m\simeq \d$ \citep{MW96}, and
%
%
%
\be
\label{bhmarkov} 
 b_h(\nu)=1+\frac{\nu^2-1}{\d_c}\, .  
\ee 

\subsection{Non-Markovian corrections}

We now use the path integral formalism discussed in Section~2.2 to
compute the non-Markovian corrections to the halo bias.
The relevant quantity for our purposes is 
the conditional probability
 \begin{eqnarray} 
&&P(\delta_n,S_n| \delta_m,S_m)\equiv\label{Pbias}\\
&=&\frac{
\int_{-\infty}^{\delta_c}d\delta_1\cdots
\widehat{d\delta_{m}}\cdots d\delta_{n-1}
W\left(\delta_0=0;\delta_1,\ldots, \delta_n;S_n\right)
}{\int_{-\infty}^{\delta_c}d\delta_1\cdots d\delta_{m-1}
W\left(\delta_0=0;\delta_1,\cdots,\delta_m;S_m\right)}\, ,\nonumber
\end{eqnarray}
where the hat over $d\delta_{m}$ means that $d\delta_{m}$ must be omitted
from the list of integration variables. The numerator is a sum over all
trajectories that start from $\d_0=0$ at $S=0$, have a given fixed value
$\d_m$ at $S_m$, and a value $\d_n$ at $S_n$, while all other points of the
trajectory, $\d_1, \ldots ,\d_{m-1},\d_{m+1},\ldots \d_{n-1}$ are
integrated from $-\infty$ to $\d_c$ .  The denominator gives the
appropriate normalization to the conditional probability.

Similarly to \eq{calFmarkov}, the conditional first-crossing rate
${\cal F}(S_n|\d_m,S_m)$ is obtained from the conditional probability
$P (\delta_n,S_n| \d_m,S_m) $ using
\be\label{firstcrossTcond}
{\cal F} (S_n|\d_m,S_m)  = -\int_{-\infty}^{\d_c}d\d_n\, 
\frac{\pa P (\delta_n,S_n| \d_m,S_m)}{\pa S_n}\, .
\ee
In the Gaussian and Markovian case, the probability $W^{\rm gm}$ satisfies
\bees\label{facto}
&&W^{\rm gm}(\delta_0;\delta_1,\ldots,\delta_n; S_n)= 
W^{\rm gm}(\delta_0;\delta_1,\ldots, \delta_m;S_m)\nn\\
&&\times W^{\rm gm}(\delta_m; \delta_{m+1}, \ldots ,\delta_n; S_n-S_m)\,,
\ees
and $P (\delta_n,S_n| \d_m,S_m) $ in \eq{Pbias} becomes identical to the
probability of arriving in $\d_n$ at time $S_n$, starting from $\d_m$ at
time $S_m$, which is given by \eq{PiChandra} (with $\d_m$ identified with
$\d_0$ at $S=S_n-S_m$), and we therefore recover the excursion set theory
result.

We have computed the non-Markovian corrections to this result for the case
of Gaussian fluctuations and a tophat filter in coordinate space, i.e. with
the two-point function given in \eqs{rever}{approxDelta}. The computation is
quite involved, and we leave the details to the Appendix. Taking finally
$S_m= 0$ and developing to first order in $\d_m\equiv \d_0$, which is the
case relevant to the computation of the bias, for the conditional first
crossing rate we find
\bees 
   &&{\cal F}(S|\d_0,S_0=0)=\frac{\d_c}{\sqrt{2\pi}\, S^{3/2}} e^{-\d_c^2/(2S)}\nn\\
   &&\times
   \left\{ \(1-\kappa+\frac{\kappa}{2} e^{\nu^2/2}\Gamma(0,\nu^2/2)
   \)\right.\label{calFfinal}\\
&&\phantom{++}\left.+\frac{\d_0}{\d_c}\[ (\nu^2-1)+\frac{\kappa}{2} \(2-
  e^{\nu^2/2}\Gamma(0,\nu^2/2)\)\]\right\}\, .\nn 
\ees
From this we obtain the Lagrangian halo bias $b_h^L$
and the Eulerian halo bias $b_h$:
\bees\label{bh}
b_h (\nu) &=& 1 + b_h^L \\  \nonumber
  &=& 1 + \frac{1}{\d_c}\,
\frac{1}{1-\kappa+\frac{\kappa}{2} e^{\nu^2/2}\Gamma(0,\nu^2/2)}\,\\
&&\times\left\{ (\nu^2-1)+\frac{\kappa}{2} \[ 2- e^{\nu^2/2}\Gamma(0,\nu^2/2) \]
\right\} \, .\nn
\ees
%
For $\kappa=0$ we recover the usual Markovian result in equation~(\ref{bhmarkov}).
For $\nu\gg1$, corresponding to large masses, $e^{\nu^2/2}\Gamma(0,\nu^2/2)\rightarrow 2/\nu^2$,
and the above expression simplifies to
\bees
\label{bhlargenu}
b_h (\nu) 
 & \rightarrow & 1 + \frac{\nu^2-1}{\d_c}\,
\(\frac{1+\frac{\kappa}{\nu^2} }{1-\kappa+\frac{\kappa}{\nu^2} } \) \nonumber\\
 & \approx & \frac{1}{1-\kappa}\, \frac{\nu^2}{\d_c}\,, \qquad {\rm for\ }
   \nu \gg 1 \, .
\ees

\begin{figure*}
   \subfigure{
   \includegraphics[width=0.45\textwidth]{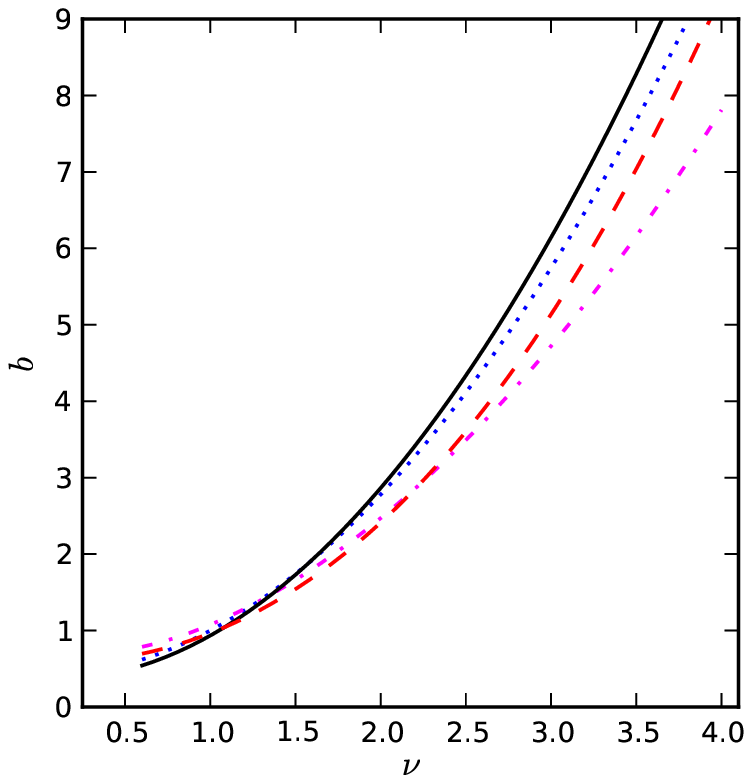} }
   \subfigure{
    \includegraphics[width=0.45\textwidth]{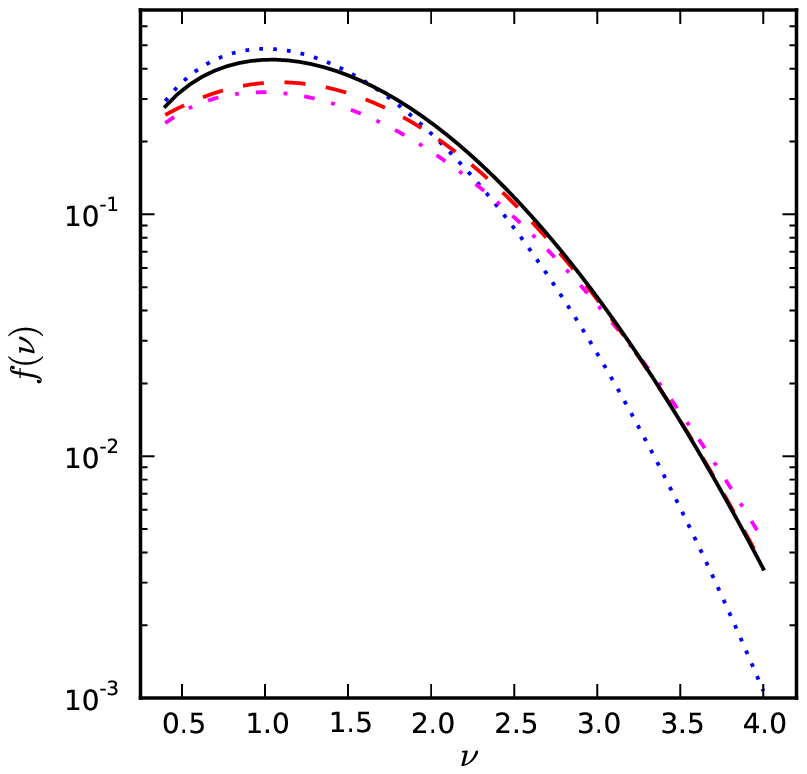} }
   \subfigure{
    \includegraphics[width=0.45\textwidth]{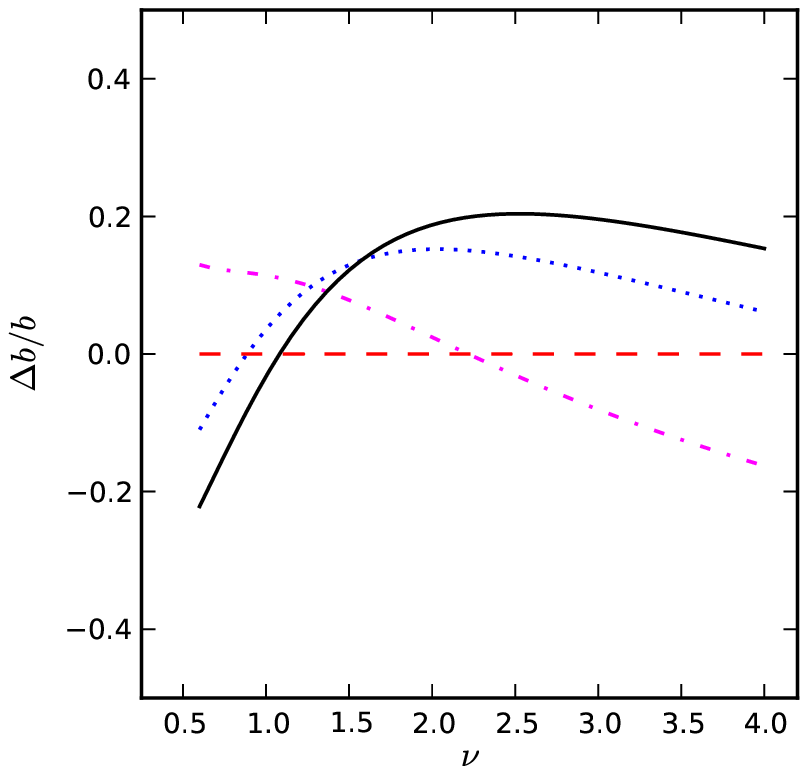} }
   \subfigure{
    \includegraphics[width=0.45\textwidth]{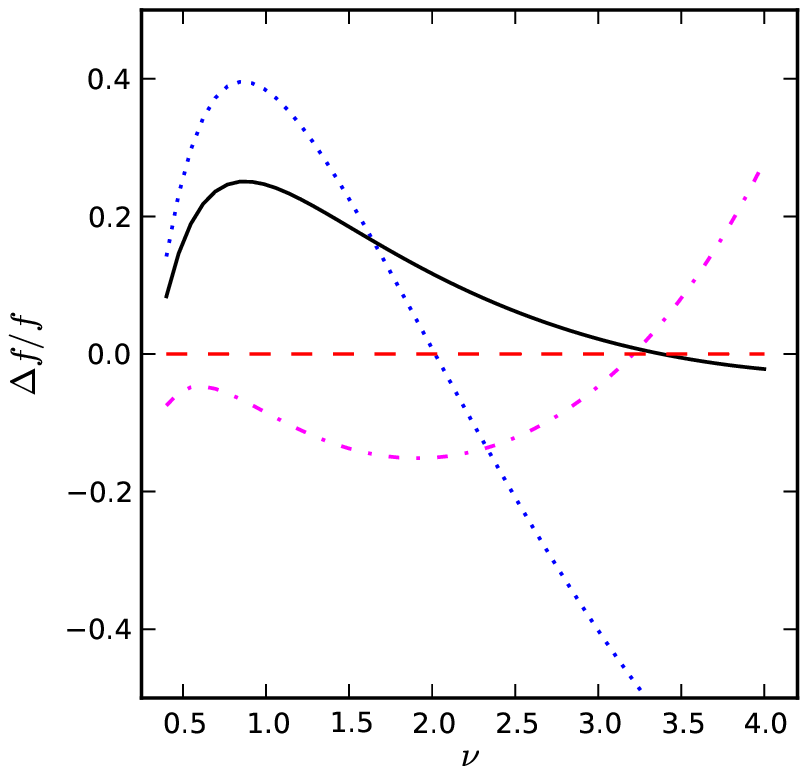} }
  \caption{ Comparison of the Eulerian halo bias $b_h(\nu)$ (left panels)
    and the halo mass function $f(\nu)$ (right panels) from various
    analytic models and simulations: our non-Markovian and stochastic
    barrier model from \eq{bhfinal}, with $a=0.818, \kappa=0.23$ (black
    solid line), the standard Markovian spherical collapse (blue dotted)
    and ellipsoidal collapse (magenta dot-dashed) models, and the fits to
    $N$-body simulation results (red dashed) from Tinker et
    al. (2008,2010).  The bottom panels show the fractional difference
    between each of the analytic model prediction and the fit to $N$-body
    result.  With only two free parameters, our new model is able to match
    the $N$-body results to within $\sim 20$\% for a wide range of $\nu$. }
\label{fig:compare_bias}
\end{figure*}



\subsection{Adding a stochastic barrier}

In the presence of the stochastic barrier described in Section~2.3, we can
easily modify the halo bias in equation~(\ref{bh}) using the substitution
$\d_c \rightarrow a^{1/2}\d_c$ and $\kappa \rightarrow a\kappa$, where the
parameter $a$ is related to the diffusion coefficient of the barrier.  The
Eulerian halo bias finally reads
\bees\label{bhfinal}
b_h(\nu)&=&1+\frac{1}{\sqrt{a} \d_c}\, 
\frac{1}{1-a\kappa+\frac{a\kappa}{2} e^{a\nu^2/2}\Gamma(0,a\nu^2/2)
}\,\\
&&\times \left\{ (a\nu^2-1)+\frac{a\kappa}{2} 
 \[ 2- e^{a\nu^2/2}\Gamma(0,a\nu^2/2) \] \right\}\, .\nn
\ees
We note that equation~(\ref{bhfinal}) raises the halo bias in the large
$\nu$ (i.e. large halo mass) region compared to the bias in the ellipsoidal
collapse model \citep{SMT01}, and in fact get closer to the spherical result.
For $\nu\gg1$, using again the asymptotic expression of the incomplete Gamma function and keeping only the leading term $\sim\nu^2$,
our result reads
\be
b_h(\nu)\simeq \frac{a^{1/2}}{(1-a\kappa)}\, \frac{\nu^2}{\d_c}\,, \qquad {\rm for\ }
   \nu \gg 1
\ee
which differs from the asymptotic spherical collapse result by an overall factor of
$a^{1/2}/(1-a\kappa)$.  

\section{Comparisons}

We now compare the predictions for the Eulerian halo bias $b_h(\nu)$ from
our non-Markovian and stochastic barrier model with those from the standard
excursion set theory as well as $N$-body simulations.  We present the
results for the halo mass function in parallel since as we have shown in
Sections~2 and 3, an analytic theory for halo formation provides simultaneous
predictions for the mass function and bias.

Our model contains two parameters: (1) $\kappa$, which parameterizes the
degree of non-Markovianity and its exact value depends on the filter
function used to smooth the density, e.g., $\kappa=0, 0.35, 0.45$ for
tophat in momentum-space, Gaussian, and tophat in coordinate filter,
respectively; (2) $a$, which parameterizes the stochasticity of the
diffusing barrier with the diffusion coefficient $D_B$, where
$a=1/(1+D_B)$.  There is no a priori reason to favor one filter to another,
nor is the choice of filters limited to the three functional forms given
above.  Furthermore, we recall that in \cite{MR1} the scaling $\kappa
\rightarrow a\,\kappa$ is obtained under the simplified assumption that the
barrier makes a simple Brownian motion around the spherical collapse
barrier; for more complicated stochastic motions of the barrier (and also
for fluctuations around the ellipsoidal barrier), the rescaling of $\kappa$
might be different.  For these reasons, we prefer in this work to treat
both $a$ and $\kappa$ as free parameters and use simulations to calibrate
their values.

For the simulations, we choose to compare with the fits to the latest
$N$-body simulations \citep{tinker08,tinker10}.  These papers provide
detailed discussions about the comparison within the different $N$-body
results and numerical issues such as the dependence of the results on
simulation resolution and halo definitions and finders.

Fig.~\ref{fig:compare_bias} shows the results for the bias (left panels)
and mass function (right panels) as a function of halo mass (as
parameterized by $\nu=\delta_c/\sigma$).  The bottom panels show the
fractional difference between each model prediction and the fit to $N$-body
results.  For the spherical and ellipsoidal models, we use the standard
parameters listed in Table~1.  For our model, we plot the predictions using
$a=0.818$ and $\kappa=0.23$, which provide a good match (within $\sim
20$\%) to both the bias and mass function from $N$-body, and in particular
to the mass function at high mass.

We note that since $\kappa$ is related to the two-point correlation
function of the density field (see eq.~91 of \cite{MR1}), it can in
principle depend on the cosmological model.  As discussed in footnote~10 of
\cite{MR1}, however, the dependence of $\kappa$ on the cosmological parameters is
extremely weak.  Both $\kappa$ and $a$ can therefore be treated as universal
parameters whose values can be calibrated with simulations.  

\section{Conclusions}

We derived an analytic expression (eq.~\ref{bhfinal} with $a=1$) for the
halo bias in the non-Markovian extension of the excursion set theory.  This
new model is based on a path integral formulation introduced in \cite{MR1},
which provides an analytic framework for handling the non-Markovian nature
of the random walk and for calculating perturbatively the non-Markovian
corrections to the standard version of the excursion set theory.  The
degree of non-Markovianity in our theory is parameterized by a single
variable, $\kappa$, whose exact value depends on the shape of the filter
function used to smooth the density field, e.g., $\kappa=0$ for a tophat
filter in momentum space, $\kappa\approx 0.35$ for a Gaussian filter, and
$\kappa \approx 0.44$ for a tophat filter in coordinate space.

As already discussed in \cite{Bond, robertson09, MR1}, changing the filter
function in the spherical collapse model from a tophat in momentum space to
a tophat in coordinate space does not help alleviate the discrepancy in the
mass function between the Press-Schechter model and $N$-body
simulations. In another word, had we plotted the corresponding curves in
Fig.~1 using $a=1$ (i.e. a constant barrier height as in the spherical
collapse model) and $\kappa=0.44$ (for a tophat filter in coordinate
space), the bias would be too high by up to $\sim 80$\% at large $\nu$
compared to the $N$-body result, and the mass function would be too low by
up to $\sim 80$\% at large mass (see also Fig.~9 of \citealt{MR1}).  Using
a Gaussian filter reduces $\kappa$ by only $\sim 20$\% and has only a minor
effect.  Additional modifications to the theory beyond including
non-Markovian corrections must therefore be introduced to match the
$N$-body results.

We have explored one such modification by allowing the barrier height
itself to be a stochastic variable (Sec.~2.3 and \citealt{MR2}).  This new
ingredient introduces a second parameter $a$ in our theory, as summarized
in Table~1.  As the solid black curves in Fig.~1 illustrate, an appropriate
choice of these two parameters for the non-Markovian correction and
stochastic barrier (e.g., $\kappa=0.23$ and $a=0.818$) produces a good
match to $N$-body results for {\it both} the halo mass function and bias,
with fractional deviations being $\sim 20$\% or less.  In comparison, the
ellipsoidal collapse model contains four fitting parameters ($a, b, c$ and
$q$; see Table~1) and does a comparable job at matching $N$-body
simulations (dot-dashed magenta curves in Fig.~1).  

Further improvement to the model presented in this paper can be obtained by
computing the bias through the excursion set theory starting from the
ellipsoidal model and including the effects of non-Markovianity. A step
towards this computation has been taken recently in \cite{DSMR}, where the
first crossing rate has been computed using the path integral method for a
generic barrier.  As the next step, one could envisage to combine the
diffusing barrier model with the ellipsoidal model, i.e. consider a barrier
that fluctuates around an average value given by the ellipsoidal collapse
model, rather than around the constant value provided by the spherical
collapse model as done in this paper. We expect this combined model to be
able to provide an even closer match to $N$-body results than the $\sim
20$\% accuracy achieved by either model alone.

\section*{Acknowledgments}

Support for CPM is provided in part by the Miller Institute for Basic
Research in Science, University of California Berkeley.  The work of MM is
supported by the Fonds National Suisse.  We thank Roman Scoccimarro for a
useful comment.

\bibliographystyle{mn2e}

\appendix 

\section{Details of the computation}

To perform the computation we use the technique discussed in detail in 
\citealt{MR1} (MR1). We consider first the numerator in \eq{Pbias}. 
We start from \eq{WnNG0}, with
the two-point function $\langle\d_i\d_j\rangle_c$ given in
\eqs{rever}{approxDelta}, and we expand
to first order in $\kappa$ (recall that $\D_{ij}$ is proportional to $\kappa$). This gives $W$ in terms of $W^{\rm gm}$,
\bees\label{WnNG}
&&W(\delta_0;\ldots ,\delta_n;S_n)=\nn\\
&&\Dl\, \,
e^{ i\sum_{i=1}^n\lambda_i\delta_i
-\frac{1}{2}\sum_{i,j=1}^n\lambda_i\lambda_j
({\rm min}(S_i,S_j) + \D_{ij})}\nn\\
&&\simeq
W^{\rm gm} (\delta_0;\ldots ,\delta_n;S_n)\nn\\
&&+\frac{1}{2}\sum_{i,j=1}^n\D_{ij}\pa_i\pa_j
W^{\rm gm} (\delta_0;\ldots ,\delta_n;S_n)\, ,
\ees
where $\D_{ij}\equiv \D(S_i,S_j)$, $\pa_i\equiv \pa/\pa\d_i$, and we have used the identity
\be
\lambda_ke^{i\sum_{j=1}^n\lambda_j\delta_j}=-i\pa_k
e^{i\sum_{j=1}^n\lambda_j\delta_j}
\ee
to transform the factor $-\D_{ij}\lambda_i\lambda_j$ coming from the expansion of the exponential
into $\D_{ij}\pa_i\pa_j$. 
It is convenient to split the sum into various pieces
 \begin{eqnarray}
\frac{1}{2}\sum_{i,j=1}^{n}\Delta_{ij}\partial_i\partial_j&=&
\frac{1}{2}\sum_{i,j=1}^{m-1}\Delta_{ij}\partial_i\partial_j+
\sum_{i=1}^{m-1}\Delta_{im}\partial_i\partial_m\nn\\
&&\hspace*{-4mm}+
\frac{1}{2}\sum_{i,j=m+1}^{n-1}\Delta_{ij}\partial_i\partial_j+
\sum_{i=m+1}^{n-1}\Delta_{in}\partial_i\partial_n\nn\\
&&\hspace*{-4mm}+\sum_{i=1}^{m-1}\Delta_{in}\partial_i\partial_n+
\Delta_{mn}\partial_m\partial_n\nn\\
&&\hspace*{-4mm}+\sum_{i=1}^{m-1}\sum_{j=m+1}^{n-1}\Delta_{ij}\partial_i\partial_j+
\sum_{j=m+1}^{n-1}\Delta_{jm}\partial_j\partial_m\, .\nn\label{A3}\\
\end{eqnarray}
Consider first the contribution from the first line of this expression. Using the factorization property (\ref{facto}) of $W^{\rm gm}$, its contribution to 
the numerator in \eq{Pbias} can be written as 
\bees
&& \int_{-\infty}^{\delta_c}d\delta_1\cdots d\d_{m-1}
d\d_{m+1}\cdots d\delta_{n-1}\nn \\
&&\[\frac{1}{2}\sum_{i,j=1}^{m-1}\Delta_{ij}\partial_i\partial_j+
\sum_{i=1}^{m-1}\Delta_{im}\partial_i\partial_m\]\nn\\
&&\hspace*{-8mm}\times W^{\rm gm}(\delta_0;\ldots, \delta_m;S_m)
W^{\rm gm}(\delta_m;  \ldots ,\delta_n; S_n-S_m)\nn\\
&=&\hspace*{-3mm}\int_{-\infty}^{\delta_c}
\hspace*{-2mm}d\delta_1\cdots d\d_{m-1}
\[\frac{1}{2}\sum_{i,j=1}^{m-1}\Delta_{ij}\partial_i\partial_j+
\sum_{i=1}^{m-1}\Delta_{im}\partial_i\partial_m\]\nn\\
&&\hspace*{-4mm}\times W^{\rm gm}(\delta_0;\ldots, \delta_m;S_m)\nn\\
&&\hspace*{-4mm}\times\int_{-\infty}^{\delta_c}d\d_{m+1}\cdots d\delta_{n-1}
W^{\rm gm}(\delta_m;  \ldots ,\delta_n; S_n-S_m)\nn \\
&+&\int_{-\infty}^{\delta_c}\hspace*{-2mm}d\delta_1\cdots d\d_{m-1}
\sum_{i=1}^{m-1}\Delta_{im}\partial_i
W^{\rm gm}(\delta_0;\ldots, \delta_m;S_m)\nn\\
&&\hspace*{-4mm}\times\int_{-\infty}^{\delta_c}
\hspace*{-2mm}d\d_{m+1}\cdots d\delta_{n-1}
\pa_mW^{\rm gm}(\delta_m;  \ldots ,\delta_n; S_n-S_m)\, .\nn
\label{A4}\\
\ees
The first term is easily dealt by observing that
\bees
&&\int_{-\infty}^{\delta_c}d\d_{m+1}\cdots d\delta_{n-1}
W^{\rm gm}(\delta_m;  \ldots ,\delta_n; S_n-S_m)\nn\\
&&=
\Pi^{\rm gm}(\d_m;\d_n;S_n-S_m)\, .
\ees
Combining this with the contribution coming from the 
zero-th order term $W^{\rm gm} (\delta_0;\ldots ,\delta_n;S_n) $ in 
\eq {WnNG} and using again the factorization property (\ref{facto}) of $W^{\rm gm}$, we therefore get
\bees
&&\Pi^{\rm gm}(\d_m;\d_n;S_n-S_m)
\int_{-\infty}^{\delta_c}d\delta_1\cdots d\d_{m-1}\nn\\
&&\times \[1+\frac{1}{2}\sum_{i,j=1}^{m-1}\Delta_{ij}\partial_i\partial_j+
\sum_{i=1}^{m-1}\Delta_{im}\partial_i\partial_m\]\nn\\
&&\times W^{\rm gm}(\delta_0;\ldots, \delta_m;S_m)\\
&+&\int_{-\infty}^{\delta_c}\hspace*{-2mm}d\delta_1\cdots d\d_{m-1}
\sum_{i=1}^{m-1}\Delta_{im}\partial_i
W^{\rm gm}(\delta_0;\ldots, \delta_m;S_m)\nn\\
&&\hspace*{-4mm}\times\int_{-\infty}^{\delta_c}
\hspace*{-2mm}d\d_{m+1}\cdots d\delta_{n-1}
\pa_mW^{\rm gm}(\delta_m;  \ldots ,\delta_n; S_n-S_m)\, .\nn
\ees
We now observe that the  term in brackets give just the expansion to $O(\kappa)$ of the denominator in \eq{Pbias}. Therefore, to $O(\kappa)$, we can write
\bees
P(\delta_n,S_n| \delta_m,S_m)&=&\Pi^{\rm gm}(\d_m;\d_n;S_n-S_m)\nn\\
&&+P^{\rm non-mark}(\delta_n,S_n| \delta_m,S_m)\, ,\label{Ptot}
\ees
where
\be\label{PPnonm}
P^{\rm non-mark}(\delta_n,S_n| \delta_m,S_m)=
\frac{N_a+N_b+N_c+N_d}{\Pi^{\rm gm}(\d_0;\d_m;S_m)}\, ,
\ee
and $N_a,\ldots ,N_d$ are defined by
\bees
N_a&=&\int_{-\infty}^{\delta_c}\hspace*{-2mm}d\delta_1\cdots d\d_{m-1}
\sum_{i=1}^{m-1}\Delta_{im}\partial_i
W^{\rm gm}(\delta_0;\ldots, \delta_m;S_m)\nn\\
&&\hspace*{-4mm}\times\int_{-\infty}^{\delta_c}
\hspace*{-2mm}d\d_{m+1}\cdots d\delta_{n-1}
\pa_mW^{\rm gm}(\delta_m;  \ldots ,\delta_n; S_n-S_m)\, ,\nn\\
\ees
\bees
N_b&=&\int_{-\infty}^{\delta_c}\hspace*{-2mm}d\delta_1\cdots d\d_{m-1}
d\d_{m+1}\cdots d\delta_{n-1}\label{Nb0}\\
&&\[
\frac{1}{2}\sum_{i,j=m+1}^{n-1}\Delta_{ij}\partial_i\partial_j+
\sum_{i=m+1}^{n-1}\Delta_{in}\partial_i\partial_n\]\nn\\
&&\hspace*{-8mm}\times W^{\rm gm}(\delta_0;\ldots, \delta_m;S_m)
W^{\rm gm}(\delta_m;  \ldots ,\delta_n; S_n-S_m)\, ,\nn
\ees
\bees
N_c&=&\int_{-\infty}^{\delta_c}\hspace*{-2mm}d\delta_1\cdots d\d_{m-1}
d\d_{m+1}\cdots d\delta_{n-1}\\
&&\[
\sum_{i=1}^{m-1}\Delta_{in}\partial_i\partial_n+
\Delta_{mn}\partial_m\partial_n
\]\nn\\
&&\hspace*{-8mm}\times W^{\rm gm}(\delta_0;\ldots, \delta_m;S_m)
W^{\rm gm}(\delta_m;  \ldots ,\delta_n; S_n-S_m)\, ,\nn
\ees
\bees
N_d&=&\int_{-\infty}^{\delta_c}\hspace*{-2mm}d\delta_1\cdots d\d_{m-1}
d\d_{m+1}\cdots d\delta_{n-1}\\
&&\[
\sum_{i=1}^{m-1}\sum_{j=m+1}^{n-1}\Delta_{ij}\partial_i\partial_j+
\sum_{j=m+1}^{n-1}\Delta_{jm}\partial_j\partial_m
\]\nn\\
&&\hspace*{-8mm}\times W^{\rm gm}(\delta_0;\ldots, \delta_m;S_m)
W^{\rm gm}(\delta_m;  \ldots ,\delta_n; S_n-S_m)\, .\nn
\ees
The contribution $N_a$ comes from \eq{A4}, while 
$N_b,N_c$ and $N_d$ come from the second, third and fourth line in
\eq{A3}, respectively.
Observe that in the denominator in \eq{PPnonm} we could replace
$\Pi(\d_0;\d_m;S_m)$ by
$\Pi^{\rm gm}(\d_0;\d_m;S_m)$, since the numerator is  proportional to $\D_{ij}$ and therefore is already ${\cal O}(\kappa)$.

The contributions $N_a,\ldots, N_d$ can be computed using the techniques developed in MR1. The term $N_a$ is immediately obtained using 
eqs.~(105) and (110) of MR1, and is given by
\bees
N_a&=& \kappa \,\frac{\d_c(\d_c-\d_m)}{S_m}
{\rm Erfc}\(\frac{2\d_c-\d_m}{\sqrt{2S_m}}\)\nn\\
&&\times
\pa_m\Pi^{\rm gm}(\d_m;\d_n;S_n-S_m)\, ,
\ees
where ${\rm Erfc}$ is the complementary error function. The term $N_b$ is given by
\bees\label{Nb}
N_b&=&\Pi^{\rm gm}(\d_0;\d_m;S_m)\\
&&\times [
\Pi^{b1} (\d_m,S_m;\d_n,S_n) +\Pi^{b2} (\d_m,S_m;\d_n,S_n)]\, ,\nn
\ees
where 
\bees
&&\Pi^{b1} (\d_m,S_m;\d_n,S_n)\equiv 
\int_{-\infty}^{\delta_c}\hspace*{-2mm}
d\d_{m+1}\cdots d\delta_{n-1}\label{Nb1}\\
&&\times
\sum_{i=m+1}^{n-1}\Delta_{in}\partial_i\partial_n
W^{\rm gm}(\delta_m;  \ldots ,\delta_n; S_n-S_m)\, ,\nn
\ees
and
\bees
&&\Pi^{b2} (\d_m,S_m;\d_n,S_n)\equiv 
\int_{-\infty}^{\delta_c}\hspace*{-2mm}
d\d_{m+1}\cdots d\delta_{n-1}\label{Nb2}\\
&&\times
\frac{1}{2}\sum_{i,j=m+1}^{n-1}\Delta_{ij}\partial_i\partial_j
W^{\rm gm}(\delta_m;  \ldots ,\delta_n; S_n-S_m)\, .\nn
\ees
The computation of $\Pi^{b1}$ and $\Pi^{b2}$ is quite similar to the computation of the terms called
$\Pi^{\rm mem}$ and $\Pi^{\rm mem-mem}$ in MR1, and in the continuum limit $\eps\ra 0$ we get
\bees\label{Pib1}
&&\hspace*{-4mm}\Pi^{b1}(\d_m,S_m;\d_n,S_n)=\pa_n\lim_{\eps\ra 0}
\frac{1}{\eps}\int_{S_m}^{S_n}dS_i\,\nn\\
&&\hspace*{-4mm}\times\Delta(S_i,S_n)\Pi_{\eps}^{\rm gm}(\d_m;\d_c;S_i-S_m)
\Pi_{\eps}^{\rm gm}(\d_c;\d_n;S_n-S_i)\nn\\
&=&\frac{\kappa}{\pi} (\d_c-\d_m) \pa_n\biggl\{(\d_c-\d_n)
\int_{S_m}^{S_n}dS_i\nn\\
&&\times \frac{S_i}{S_n(S_i-S_m)^{3/2}(S_n-S_i)^{1/2}}\nn\\
&&\times
\exp\[-\frac{(\d_c-\d_m)^2}{2(S_i-S_m)}-\frac{(\d_c-\d_n)^2}{2(S_n-S_i)}\]\biggr\}\,
\label{finalb1}
\ees
and 
\bees
&&\hspace*{-4mm}\Pi^{b2}(\d_m,S_m;\d_n,S_n)=\lim_{\eps\ra 0}
\frac{1}{\eps^2}\int_{S_m}^{S_n}dS_i\int_{S_i}^{S_n}dS_j\,\nn\\
&&\hspace*{-4mm}\times \Delta(S_i,S_j)\Pi_{\eps}^{\rm gm}(\d_m;\d_c;S_i-S_m)\nn\\
&&\hspace*{-4mm}\times \Pi_{\eps}^{\rm gm}(\d_c;\d_c;S_j-S_i)
\Pi_{\eps}^{\rm gm}(\d_c;\d_n;S_n-S_j)\nn\\
&=&\frac{\kappa}{\pi\sqrt{2\pi}}\, (\delta_c-\d_m)(\delta_c-\delta_n)\nn\\
&&\hspace*{-4mm}\times
\int_{S_m}^{S_n}dS_i\,  \frac{S_i}{(S_i-S_m)^{3/2}}e^{-(\delta_c-\d_m)^2/[2(S_i-S_m)]}\nn\\
&&\hspace*{-4mm}\times\int_{S_i}^{S_n}dS_j
\frac{e^{-(\delta_c-\delta_n)^2/[2(S_n-S_j)]}}{S_j(S_j-S_i)^{1/2}
(S_n-S_j)^{3/2} }
\, .
\ees
This can be rewritten as a total derivative with respect to $\d_n$, as
\bees
&&\hspace*{-4mm}\Pi^{b2}(\d_m,S_m;\d_n,S_n)=
\frac{\kappa}{\pi\sqrt{2\pi}}\, (\delta_c-\d_m)\pa_n\nn\\
&&\hspace*{-4mm}\times
\int_{S_m}^{S_n}dS_i\,  \frac{S_i}{(S_i-S_m)^{3/2}}e^{-(\delta_c-\d_m)^2/[2(S_i-S_m)]}\nn\\
&&\hspace*{-4mm}\times\int_{S_i}^{S_n}dS_j
\frac{e^{-(\delta_c-\delta_n)^2/[2(S_n-S_j)]}}{S_j(S_j-S_i)^{1/2}
(S_n-S_j)^{1/2} }
\, .\label{finalb2}
\ees
The fact that both $\Pi^{b1}$ and $\Pi^{b2}$ can be written as a derivative with respect to $\d_n$ simplifies considerably the computation of the flux ${\cal F}(S)$, since we can integrate $\pa_n\equiv\pa/\pa\d_n$ by parts, and then we only need to evaluate the integrals  in \eqs{finalb1}{finalb2} in $\d_n=\d_c$, which can be done analytically, as discussed in MR1.

The term $N_c$ is a total derivative with respect to $\pa_n$ of a quantity that vanishes in $\d_n=\d_c$ so, when inserted into \eq{firstcrossTcond}, it gives a vanishing contribution to the first crossing rate. The most complicated term is $N_d$. Using the techniques developed in MR1, a rather long computation gives
\bees
N_d&=&\frac{\kappa}{\pi}\pa_n\Bigl\{
\d_c (\d_c-\d_m) {\rm Erfc}\(\frac{2\d_c-\d_m}{\sqrt{2S_m}}\)\pa_m I\label{Nd}\\
&&\phantom{\frac{\kappa}{\pi}\pa_n}
+  \Pi^{\rm gm}(\d_0;\d_m;S_m)\tilde{N}_d\Bigr\}\, ,
\ees
where
\bees
\tilde{N}_d&=&
-\d_m(\d_c-\d_m) I (\d_m,\d_n)+S_m (\d_c-\d_m)\pa_m I (\d_m,\d_n) 
\nn\\
&&-S_m I (\d_m,\d_n)\, ,
\ees
and
\bees
I (\d_m,\d_n)&\equiv &\int_{S_m}^{S_n}dS_j
\frac{1}{S_j (S_j-S_m)^{1/2}(S_n-S_j)^{1/2}}\nn\\
&&\hspace*{-3mm}\times \exp\left\{-\frac{(\d_c-\d_m)^2}{2 (S_j-S_m)}
-\frac{(\d_c-\d_n)^2}{2 (S_n-S_j)}\right\}\, .
\ees
Using \eqs{PPnonm}{Nb}, \eq{Ptot} can be rewritten as
\be
P(\delta_n,S_n| \delta_m,S_m)=\Pi^{\rm gm}+\Pi^{b1}+
\Pi^{b2}+\frac{N_a+N_c+N_d}{\Pi^{\rm gm}(\d_0;\d_m;S_m)}\, .
\ee
We can now compute the contribution to the flux from the various terms. 
The term $\Pi^{\rm gm}$ gives the  zero-th order term,
\bees
&&\hspace*{-3mm}{\cal F}^{\rm gm}(S_n|\d_m,S_m)=-\frac{\pa}{\pa S_n}\int_{-\infty}^{\d_c} 
d\d_n\, 
\Pi^{\rm gm} (\d_m;\d_n;S_n-S_m)\nn\\
&&\hspace*{-3mm}=\frac{1}{\sqrt{2\pi}}\, \frac{\d_c-\d_m}{(S_n-S_m)^{3/2}}
e^{-(\d_c-\d_m)^2/[2(S_n-S_m)]}\, .
\ees
The contribution of $\Pi^{b1}$ to the flux is zero since it is the derivative with respect to $\pa_n$ of a quantity that vanishes in $\d_n=\d_c$, and the same holds for $N_c$.
The contribution of $\Pi^{b2}$   is
\bees
&&{\cal F}^{b2}(S_n|\d_m,S_m)\\
&&=
-\frac{\pa}{\pa S_n}\int_{-\infty}^{\d_c} d\d_n\, 
\Pi^{b2} (\d_m,S_m;\d_n;S_n)\nn\\
&&=
-\frac{\kappa}{\pi\sqrt{2\pi}}\, (\delta_c-\d_m)\frac{\pa}{\pa S_n}\nn\\
&&\hspace*{-4mm}\times\int_{S_m}^{S_n}dS_i\,  \frac{S_i}{(S_i-S_m)^{3/2}}
e^{-(\delta_c-\d_m)^2/[2(S_i-S_m)]}\nn\\
&&\hspace*{-4mm}\times\int_{S_i}^{S_n}dS_j
\frac{1}{S_j(S_j-S_i)^{1/2}
(S_n-S_j)^{1/2} }
\, .
\ees
The inner integral is elementary,
\be
\int_{S_i}^{S_n}dS_j
\frac{1}{S_j(S_j-S_i)^{1/2}(S_n-S_j)^{1/2} }=\frac{\pi}{(S_iS_n)^{1/2}}\, ,
\ee
and we end up with
\bees
&&{\cal F}^{b2}(S_n|\d_m,S_m)=
-\frac{\pa}{\pa S_n}\Bigl[ 
\frac{\kappa (\delta_c-\d_m)}{\sqrt{2\pi S_n}}\\
&&\times
\int_{S_m}^{S_n}dS_i\,  \frac{S^{1/2}_i}{(S_i-S_m)^{3/2}}
e^{-(\delta_c-\d_m)^2/[2(S_i-S_m)]}
\Bigr]\, ,\nn
\ees
which generalized eq.~(118) of MR1 to $\d_m\neq 0$ and $S_m\neq 0$.
For $S_m$  generic the integral cannot be performed analytically. However, for computing the bias we are actually interested in the limit $S_m\ra 0$ with $\d_m$ generic, and we see that in this case this contribution reduces to that computed in MR1, with the replacement 
$\d_c\ra\d_c-\d_m$.

The remaining contributions can be computed similarly. For the term $N_d$, 
again, rather than computing explicitly the derivative $\pa_n=\pa/\pa\d_n$
in \eq{Nd}, it is convenient to insert this expression directly into the first-crossing rate 
(\ref{firstcrossTcond}), and use the fact that it is a total derivative with respect to $\pa_n$ to perform the integral over $d\d_n$.  So, in the end, we only need
\bees
I (\d_m,\d_n=\d_c)
&=&\frac{\pi}{(S_mS_n)^{1/2}} e^{+(\d_c-\d_m)^2/(2 S_m)}\label{Idndc}\\
&&\times
{\rm Erfc}\[(\d_c-\d_m)\sqrt{\frac{S_n}{2S_m(S_n-S_m)}}\]\, .\nn
\ees
We can now put together all the terms and   take the limit $S_m\ra 0$ (with $\d_c-\d_m>0$).
In this limit the ${\rm Erfc}$ function in \eq{Idndc} reduces to an exponential, so
\be
I (\d_m,\d_n=\d_c)\simeq \frac{\sqrt{2\pi (S_n-S_m)}}{(\d_c-\d_n)S_n}\, 
e^{-(\d_c-\d_n)^2/[2(S_n-S_m)]}\, .
\ee
Denoting $\d_m=\d_0$ in this limit, we finally get 
\bees
{\cal F} (S|\d_0,S_m=0)  &=&\frac{1-\kappa}{\sqrt{2\pi}}
\,\frac{\d_c-\d_0}{S^{3/2}} e^{-(\d_c-\d_0)^2/(2S)}\nn\\
&&\hspace*{-15mm}+\frac{\kappa}{2\sqrt{2\pi}}\,\frac{\d_c-\d_0}{S^{3/2}} 
\Gamma\(0,\frac{(\d_c-\d_0)^2}{2S}\)\\
&&\hspace*{-15mm}-\frac{\kappa}{\sqrt{2\pi}}\, \frac{\d_0}{S^{3/2}}
\[ 1- \frac{(\d_c-\d_0)^2}{S}\]  e^{-(\d_c-\d_0)^2/(2S)}\, .\nn
\ees
Expanding this result to first order in $\d_0$ we obtain 
\eq{calFfinal}.

\label{lastpage}

\end{document}

%% file: mydefs.tex
\renewcommand\({\left(}
\renewcommand\){\right)}
\renewcommand\[{\left[}
\renewcommand\]{\right]}

\newcommand{\ra}{\rightarrow}

\def\lsim{\raise 0.4ex\hbox{$<$}\kern -0.8em\lower 0.62
ex\hbox{$\sim$}}

\def\gsim{\raise 0.4ex\hbox{$>$}\kern -0.7em\lower 0.62
ex\hbox{$\sim$}}

\def\lbar{{\hbox{$\lambda$}\kern -0.7em\raise 0.6ex
\hbox{$-$}}}

\newcommand\eq[1]{eq.~(\ref{#1})}
\newcommand\eqs[2]{eqs.~(\ref{#1}) and (\ref{#2})}

\newcommand\pa{\partial}
\newcommand\p{\partial}

\newcommand\ee{\end{equation}}
\newcommand\be{\begin{equation}}
\def\bea{\begin{array}}
\def\eea{\end{array}}\def\ea{\end{array}}
\newcommand\ees{\end{eqnarray}}
\newcommand\bees{\begin{eqnarray}}

\def\p1{{\bf p}_1}
\def\p2{{\bf p}_2}
\def\k1{{\bf k}_1}
\def\k2{{\bf k}_2}

\newcommand{\dddM}{\kern 0.2em \raise 1.9ex\hbox{$...$}\kern -1.0em \hbox{$M$}}
\newcommand{\dddQ}{\kern 0.2em \raise 1.9ex\hbox{$...$}\kern -1.0em \hbox{$Q$}}
\newcommand{\dddI}{\kern 0.2em \raise 1.9ex\hbox{$...$}\kern -1.0em\hbox{$I$}}
\newcommand{\dddJ}{\kern 0.2em \raise 1.9ex\hbox{$...$}\kern-1.0em
\hbox{$J$}}
\newcommand{\dddcalJ}{\kern 0.2em \raise 1.9ex\hbox{$...$}\kern-1.0em
\hbox{${\cal J}$}}

\newcommand{\dddO}{\kern 0.2em \raise 1.9ex\hbox{$...$}\kern -1.0em
\hbox{${\cal O}$}}
\def\dddz{\raise 1.5ex\hbox{$...$}\kern -0.8em \hbox{$z$}}
\def\dddd{\raise 1.8ex\hbox{$...$}\kern -0.8em \hbox{$d$}}
\def\dddbd{\raise 1.8ex\hbox{$...$}\kern -0.8em \hbox{${\bf d}$}}
\def\ddbd{\raise 1.8ex\hbox{$..$}\kern -0.8em \hbox{${\bf d}$}}
\def\dddx{\raise 1.6ex\hbox{$...$}\kern -0.8em \hbox{$x$}}

\def\D{\Delta}
\def\p{\partial}

\def\nn{\nonumber}

\def\s{\sigma}

\def\G{\Gamma}
\def\d{\delta}

\def\eps{\epsilon}

\def\dslash{\hspace{-1mm}\not{\hbox{\kern-2pt $\partial$}}}
\def\Dslash{\not{\hbox{\kern-4pt $D$}}}
\def\pslash{\not{\hbox{\kern-2.1pt $p$}}}
\def\kslash{\not{\hbox{\kern-2.3pt $k$}}}
\def\qslash{\not{\hbox{\kern-2.3pt $q$}}}

\newcommand{\inT}{\int_{-\infty}^{\infty}}

\newcommand{\Dl}{\int{\cal D}\lambda}

%% file: bias.bbl
\begin{thebibliography}
{}

\bibitem[Bardeen et al.(1986)]{Bardeen}
Bardeen J.M., Bond J.R., Kaiser N.  and Szalay A., 1986, ApJ  304, 15.

\bibitem[Bond et al.(1991)]{Bond} Bond J., Cole S., Efstathiou G., Kaiser~N., 1991, ApJ, 379, 440.

\bibitem[Cole \& Kaiser(1989)]{CK89} Cole S. \& Kaiser~N., 1989,
MNRAS, 237, 1127.

\bibitem[De Simone et al.(2010)]{DSMR}
De Simone A., Maggiore M. and Riotto A., 
arXiv:1007.1903 [astro-ph.CO], submitted to MNRAS.

\bibitem[Efstathiou et al.(1988)]{EFWD}  Efstathiou G., Frenk~C.~S., 
White~S.~D.~M., Davis,~M., 1988, MNRAS 235, 715.

\bibitem[Jing(1998)]{jing98} Jing Y. P., 1998, ApJ, 503, L9

\bibitem[Kaiser(1984)]{Kaiser84} Kaiser N., 1984, ApJL, 284, L9

\bibitem[Maggiore \& Riotto(2010a)]{MR1} Maggiore, M.  \& Riotto, A., 2010, ApJ, 711, 907.

\bibitem[Maggiore \& Riotto(2010b)]{MR2} Maggiore, M.  \& Riotto, A., 2010, ApJ, 717, 515

\bibitem[Maggiore \& Riotto(2010c)]{MR3} Maggiore, M.  \& Riotto, A., 2010, ApJ, 717, 526

\bibitem[Maggiore \& Riotto(2010d)]{MR4} Maggiore, M.  \& Riotto, A., 2010, MNRAS, 405, 1244

\bibitem[Mo \& White(1996)]{MW96} Mo H. \& White S., 1996, MNRAS, 282, 347.

\bibitem[Pillepich et al(2010)]{pillepich10}
Pillepich A., Porciani, C., Hahn, O., 2010, MNRAS, 402, 191

\bibitem[Press \& Schechter(1974)]{PS} 
Press W. H. \& Schechter P., 1974, ApJ, 187, 425.

\bibitem[Robertson et al.(2009)]{robertson09} Robertson, B., Kravtsov, A., Tinker, J., Zentner, A., 2009, ApJ, 696, 636

\bibitem[Seljak \& Warren(2004)]{SW04} Seljak U., Warren M., 2004, MNRAS, 355, 129

\bibitem[Sheth et al(2001)]{SMT01} Sheth R., Mo H., Tormen G., 2001, MNRAS, 323, 1
	
\bibitem[Sheth \& Tormen(1999)]{ST99} Sheth R. \& Tormen G., 1999, MNRAS, 308, 119
	
\bibitem[Sheth \& Tormen(2002)]{ST02} Sheth R. \& Tormen G., 2002, MNRAS, 329, 61 

\bibitem[Tinker	et al(2005)]{tinker05} Tinker J., Weinberg, D., Zheng Z., Zehavi I., 2005, ApJ, 631, 41

\bibitem[Tinker	et al(2008)]{tinker08} Tinker J. et al. 2008, ApJ, 688, 709

\bibitem[Tinker	et al(2010)]{tinker10} Tinker J. et al. 2010, arXiv:1001.3162

\bibitem[Zentner(2007)]{zentner07} Zentner A., 2007, Int. J. Mod. Phys. D, 16, 763.
	

\end{thebibliography}
